\documentclass[manuscript]{aastex}
\usepackage{txfonts}

\shorttitle{Solar-Cycle 24 Hemispheric Helicity Trend} \shortauthors{Hao \& Zhang}

\usepackage{mathrsfs}
\usepackage{graphicx}

\begin{document}

\title{Hemispheric Helicity Trend for Solar Cycle 24}

\author{Juan Hao and Mei Zhang}
\affil{Key Laboratory of Solar Activity, National Astronomical Observatory, Chinese Academy of Sciences, A20 Datun Road, Chaoyang District, Beijing 100012, China} \email{haojuan@nao.cas.cn}

\begin{abstract}
Using vector magnetograms obtained with the Spectro-polarimeter (SP) on aboard {\it Hinode} satellite, we studied two helicity parameters (local twist and current helicity) of 64 active regions occurred in the descending phase of solar cycle 23 and the ascending phase of solar cycle 24. Our analysis gives the following results. (1) The 34 active regions of the solar cycle 24 follow the so-called hemispheric helicity rule, whereas the 30 active regions of the solar cycle 23 do not. (2) When combining all 64 active regions as one sample, they follow the hemispheric helicity sign rule as in most other observations. (3) Despite with the so-far most accurate measurement of vector magnetic field given by SP/Hinode, the rule is still weak with large scatters. (4) The data show evidence of different helicity signs between strong and weak fields, confirming previous result from a large sample of ground-based observations. (5) With two example sunspots we show that the helicity parameters change sign from the inner umbra to the outer penumbra, where the sign of penumbra agrees with the sign of the active region as a whole. From these results, we speculate that both the $\Sigma$-effect (turbulent convection) and the dynamo have contributed in the generation of helicity, whereas in both cases turbulence in the convection zone has played a significant role.
\end{abstract}



\section{Introduction}

The concept of magnetic helicity was introduced to solar physics in the 1980s (Heyvaerts \& Priest 1984; Berger \& Field 1984) and has attracted great attentions since that. It is a physical quantity that measures the topological complexity of magnetic field such as the degree of linkage or twistedness in the field (Moffatt 1985; Berger \& Field 1984) and has been considered important in modeling many solar phenomenon such as coronal mass ejections (Zhang \& Low 2005; Zhang et al. 2006; Zhang \& Flyer 2008). The helicity of magnetic fields may be characterized by several different parameters (Moffatt, 1978) such as magnetic helicity ($H_{m}$) and current helicity ($H_{c}$). However, only the vertical component of current helicity density $h_{c}$ (and the local twist $\alpha$ etc.) can be practically computed by using vector magnetograms.

Seehafer (1990) was the first to statistically study the sign of magnetic helicity of solar active regions using magnetograms. He
estimated current helicity $h_{c}$ of 16 active regions by using extrapolation of measured photospheric magnetic fields and
concluded that in active regions the current helicity is predominantly negative in the northern hemisphere and positive in the southern
hemisphere. This tendency is the so-called ``hemispheric helicity sign rule". In the following two decades, many researchers (Pevtsov
et al. 1995, 2001, 2008; Abramenko et al. 1997; Bao \& Zhang 1998; Hagino \& Sakurai 2004; Zhang 2006) have studied and confirmed this rule by using data sets obtained with different instruments located in different places of the world, e.g. the University of Hawaii Haleakala Stokes Polarimeter (HSP) and The Marshall Space Flight Center (MSFC) Vector Magnetogaph in the US, the Solar Magnetic Field Telescope (SMFT) in China, the Mitaka Solar Flare Telescope (SFT) and The Okayama Astrophysical Observatory Solar Telescope (OAO) in Japan. It is believed that the usual hemispheric helicity sign rule is there for all three solar cycles observed (that is, solar cycles 21, 22, 23).

However, there are also some debates on this rule. For instance, Bao et al. (2000) found that $h_{c}$ in their data showed an
opposite hemispheric preference at the beginning of solar cycle 23. Hagino \& Sakurai (2005) also reported that the hemispheric helicity sign rule may not be satisfied in the solar minimum phase. Choudhuri et al. (2004) developed a model that predicts deviations from the usual hemispheric rule at the beginning of a solar cycle. However, Pevtsov et al. (2001) argued that the usual hemispheric helicity sign rule still holds for the first four years of solar cycle 23 although by nature it is a weak rule with significant scatter. Pevtsov et al. (2008) further compared data from four different instruments and concluded that ``the notion that the hemispheric helicity rule changes sign in some phases of solar cycle is not supported at a high level of significance".

Apart from these arguments, Zhang (2006) did a statistical study using 17,200 vector magnetograms obtained by SMFT. She separated her data into two parts, the weak fields (100 G $< |B_z| <$ 500 G) and the strong fields ($|B_z| > $1000 G). She calculated the $\alpha$ and $h_{c}$ of weak and strong fields separately and found that the weak magnetic fields follow the usual hemispheric helicity sign rule but strong fields not. She interpreted this as the reason why Bao et al. (2000) found the $h_{c}$ in their data violates the usual hemispheric helicity sign rule whereas $\alpha$ not.

Since its launch in September 2006, Hinode has provided us with high spatial-resolution vector magnetograms for both the descending phase of solar cycle 23 and the ascending phase of solar cycle 24. This gives us a unique chance in this Letter to use these so-far most accurate vector magnetic field measurements to shed a light on above arguments. We organize our paper as follows. In section 2, we describe the observations and data reduction. In section 3, we present our analysis and results. We conclude with a discussion in the last section.


\section{Observation and Data Reduction}

We used vector manetiograms obtained by the Spectro-polarimeter (SP) aboard Hinode (Kosugi et al. 2007). SP/Hinode obtains line profiles of two magnetically sensitive Fe lines at 630.15 and 630.25 nm and nearby continuum, using a $0.16''\times164''$ slit. There are four mapping mode of operation: Normal Map, Fast Map, Dynamics and Deep Magnetogram (Tsuneta et al. 2008). In this study we only use the normal maps and fast maps. The resolution of these magnetograms is about 0.32$''$/pixel for fast maps and 0.16$''$/pixel for normal maps.

For the period we studied, that is, from November 2006 to September 2010, there are totally 190 active regions (ARs) appeared on the Sun, that is, from NOAA 10921 to NOAA 11110. However, not every active region has been observed by SP/Hinode. We searched those active regions observed by SP/Hinode using following criteria: (1) If more than one magnetograms have been obtained for the same active region, then we only use the one that is most close to the disk center. (2) Both the longitude and latitude of the active region when observed are within $40\arcdeg$ from the disk center. This gives a total number of 64 active regions to form the sample, including 30 active regions in solar cycle 23 and 34 active regions in solar cycle 24.

The SP data are calibrated and inverted at the Community Spectro-polarimetric Analysis Ceneter (CSAC, http://www.csac.hao.ucar.edu/). The inversion is based on the assumption of the Milne-Eddington atmosphere model and a nonlinear least-square fitting technique is used to fit analytical Stokes profiles to the observed profiles. The inversion gives 36 parameters including the three components of magnetic field (field strength $B$, field inclination $\gamma$ and field azimuth $\phi$), the stray light fraction ($1-f$, where $f$ is the filling factor), and so on. The 180$\arcdeg$ azimuth ambiguity was resolved by setting the directions of the transverse fields most closely to a current-free field, an approach that was used in most other studies.

We calculated two different helicity parameters, $\alpha_{z}$ and $\alpha_{hc}$, for these 64 ARs. $\alpha_{z}$ is the mean value of local twist, defined as
\begin{equation}
\alpha_{z}=\overline{(\bigtriangledown \times \textbf{\emph{B}})_{z}/B_{z}}~.
\end{equation}
$\alpha_{hc}$ is the normalized mean current helicity density, obtained by
\begin{equation}
\alpha_{hc}= \frac{\sum (\bigtriangledown \times \textbf{\emph{B}})_{z} B_{z}}{\sum B_{z}^{2}}~.
\end{equation}
Both the averaging and integral are done over the whole magnetogram.  The definition here gives the parameter $\alpha_{hc}$ the same unit of $\alpha_{z}$, and is same to the $\alpha_{g}$ parameter discussed in Tiwari et al. (2009). In our calculation we have only used points whose total wavelength-integrated polarization is larger than $10^{-2}$, which is about three times of the polarization noise level (Lites et al. 2008). This is a criteria applied to all helicity parameter calculations, upon all other criteria we apply in following analysis.

In calculating $\alpha_{z}$ and $\alpha_{hc}$, we have used two different representations of magnetic field measurement. One is related to ``flux density", where the longitudinal magnetic field $B_z = f \cdot B \cos(\gamma)$ and the transverse magnetic field $B_t = \sqrt{f} \cdot B \sin(\gamma)$. The other is the ``field strength" where $B_z = B \cos(\gamma)$ and $B_t = B \sin(\gamma)$. Hereafter we present the first type as $B_z^1$, $B_t^1$ and the second type as $B_z^2$, $B_t^2$.  Correspondingly helicity parameters are also hereafter presented as  $\alpha^1_{z}$, $\alpha^1_{hc}$ and $\alpha^2_{z}$, $\alpha^2_{hc}$ respectively. In most previous studies researchers used the helicity parameters of the first type, that is, based on the flux density measurement of magnetic field. Due to the precise measurement of SP on board {\it Hinode}, an accurate measurement of filling factor and hence of field strength becomes possible. Thus in this Letter we calculate the helicity parameter of the second type too, in order to check whether our results depend on the type of magnetic field measurement or not.


\section{Analysis and Results}

Figure 1 presents the variation of $\alpha_{z}^1$ (left panels) and $\alpha_{z}^2$ (right panels) with the solar latitude for the 30 ARs in the descending phase of solar cycle 23 (top panels), the 34 ARs in the ascending phase of solar cycle 24 (middle panels) and the total 64 ARs (bottom panels). Here $\alpha_{z}^1$ and $\alpha_{z}^2$ are calculated only using points with $|B_{z}^1| > $ 100 G or $|B_{z}^2| > $ 100 G . The solid lines indicate the results of least-square linear fits.

Similarly, Figure 2 gives the the variation of $\alpha_{hc}^1$ (left panels) and $\alpha_{hc}^2$ (right panels) with the solar latitude for the 30 ARs in solar cycle 23 (top panels), the 34 ARs in solar cycle 24 (middle panels) and the total 64 ARs (bottom panels). The $\alpha_{hc}^1$ and $\alpha_{hc}^2$ are also calculated only using points with $|B_{z}^1| > $ 100 G or $|B_{z}^2| > $ 100 G.

Values of $d\alpha/d\theta$ from the linear fittings are also shown in Figures 1 and 2, in the unit of $10^{-9}m^{-1}deg^{-1}$. Here we see that for the 30 ARs of solar cycle 23, $d\alpha/d\theta$ for $\alpha_{z}^1$, $\alpha_{z}^2$, $\alpha_{hc}^1$ and $\alpha_{hc}^2$ are all positive. Out of these 30 ARs, only 8 $(27 \%)$ ARs of the $\alpha^1_{z}$ and 14 $(47 \%)$ ARs of the $\alpha^1_{hc}$ obey the usual hemisphere sign rule. This means that ARs in the descending phase of solar cycle 23 do not follow the usual hemispheric helicity sign rule. This is consistent with Tiwari et al. (2009) who made a similar conclusion from a sample combining data from three instruments.

Contrary to that in solar cycle 23, for the 34 ARs of solar cycle 24, 20 $(59 \%)$ ARs of the $\alpha^1_{z}$ and 20 $(59 \%)$ ARs of the $\alpha^1_{hc}$ obey the usual hemisphere sign rule. $d\alpha/d\theta$ for $\alpha_{z}^1$, $\alpha_{z}^2$, $\alpha_{hc}^1$ and $\alpha_{hc}^2$ are all negative. This means that ARs in the ascending phase of solar cycle 24 follow the usual hemispheric helicity sign rule, contrary to the prediction made in Choudhuri et al. (2004). Note that ARs in the descending phase of solar cycle 23 do show a deviation from the usual hemispheric helicity sign rule. We speculate that the physical process described in Choudhuri et al. (2004), that is, poloidal flux lines getting wrapped around a toroidal flux tube rising through the convection zone to give rise to the helicity, may still apply, but a phase shift may be required in the dynamo model used.

For all of the 64 ARs, 28 $(44 \%)$ ARs of the $\alpha^1_{z}$ and 34 $(53 \%)$ ARs of the $\alpha^1_{hc}$ follow the usual hemisphere sign rule. As a whole, these 64 ARs still follow the usual hemispheric helicity sign rule, with $d\alpha/d\theta$ for $\alpha_{z}^1$, $\alpha_{z}^2$, $\alpha_{hc}^1$ and $\alpha_{hc}^2$ all negative. This is consistent with the results from most previous studies, that is, most ARs follow the usual hemispheric helicity sign rule.

An interesting observation is that, despite for the fact that we have used the so-far most accurate measurement of vector magnetic field given by SP/Hinode, the hemispheric helicity sign rule, either indicated by the 34 ARs in solar cycle 24 or by the 64 ARs as a whole, is still weak with large scatters. As an evidence, we see from Figures 1 and 2 that the magnitudes of the correlation coefficients between the latitude and the helicity parameters are all low, with the maximum magnitude only being 0.21. This seems indicating that the large scatter is an inherent property of the rule, not caused by the measurement errors. This is consistent with the prediction in Longcope et al. (1998), where helicity is considered to be produced in the process of magnetic flux tubes rising through the solar convection zone and being buffeted by turbulence with a non-vanishing kinetic helicity ($\Sigma-$ effect).

When calculating the $\alpha_{z}^1$, $\alpha_{z}^2$, $\alpha_{hc}^1$ and $\alpha_{hc}^2$ in Figures 1 and 2 we have only used points with $|B_{z}^1| > $ 100 G or $|B_{z}^2| > $ 100 G. Now we went further to calculate $\alpha_z^1$, $\alpha_z^2$, $\alpha_{hc}^1$ and $\alpha_{hc}^2$ for $|B_{z}^1|$ or $|B_z^2| > $ 200, 300, 400 G and so on until for $|B_{z}^1|$ or $|B_z^2| >$ 2000 G. This not only allows us to check whether our results depend on the selection of $|B_z|$ threshold, but also gives us a chance to examine how the hemispheric helicity sign rule might vary with the increase of field strength.

Results of the obtained $d\alpha/d\theta$ with different $|B_z|$ thresholds are plotted in Figure 3 for all four helicity parameters. We see here that when changing the $|B_z|$ threshold from 100 G to 200G or even to 500 G, the sign of $d\alpha/d\theta$ does not change. This suggests that our above conclusion, that is, ARs in the descending phase of solar cycle 23 does not follow the usual hemispheric helicity sign rule and the ARs in the ascending phase of solar cycle 24 do, is not very sensitive to the $|B_z|$ threshold we selected.

At the same time, we see from the middle and bottom panels of Figure 3 that, when the $|B_z|$ threshold goes to high values such as 1200 G for $\alpha_{z}^1$ and $\alpha_{hc}^1$ or 1800 G for $\alpha_{z}^2$ and $\alpha_{hc}^2$, the sign of $d\alpha/d\theta$ becomes opposite to those with $|B_z|$ low. This suggests that the weak and strong fields have opposite helicity sign, as first pointed out in Zhang (2006) and later confirmed by Su et al. (2009).

Zhang (2006) only points out that strong and weak fields have opposite helicity sign, here we use two examples to show that this actually presents that on average sunspot umbra and penumbra show opposite helicity sign.  In Figure 4 we show the continuum image (top left panel) and the electric current map (top right panel) of NOAA 10940, appeared on Feb. 1, 2007 of the solar cycle 23. Circles in these two images show where the distance to the sunspot center ($r$) are $5\arcsec$, $10\arcsec$, $15\arcsec$ and $20\arcsec$ respectively. The middle right panel shows the variation of electric current $J_{z}^1=\mu_0 (\bigtriangledown \times \textbf{\emph{B}})_{z}^1$ with $r$, where the dots give the values of $J_{z}^1$ and the double-shelled curve shows the mean value of $J_{z}^1$ with a bin of 1$\arcsec$ in $r$. Similarly, the bottom left and right panels respectively show the $h_{c}^1=J_z^1 B_{z}^1$ and $\alpha_{z}^1$ values and their averages with a bin of 1$''$ in $r$. We see clearly here that the inner most fields (where $r < 5 ''$) have a positive average value of $h_c^1$ or $\alpha_z^1 $ and the average value becomes negative when $r > 5 ''$.

The mean value of $|B_z|$ in the central umbra ($r\leq5\arcsec) $ is 2976 G, and is 970 G for fields in $5\arcsec<r\leq20\arcsec $.
The mean value of $\alpha_{z}^1$ in $r\leq5\arcsec$ is 5.033$\times10^{-8}m^{-1}$ and is -0.717$\times10^{-8}m^{-1}$ for regions in
$5\arcsec<r\leq20\arcsec$. The mean value of $h_{c}^1$ in $r\leq5\arcsec$ is 3.942$\times10^{-2}G^{2}m^{-1}$ and is -0.543$\times 10^{-2}G^{2}m^{-1}$ for regions in $5\arcsec<r\leq20\arcsec$. For the whole active region, with $|B_{z}| > $ 100 G, $\alpha^1_{z}=-3.274\times 10^{-9}m^{-1}$ and $\alpha^1_{hc}=-1.332\times 10^{-8}m^{-1}$. This means that the sign of the whole AR is dominated by the sign of weak field (penumbra), as also pointed out in Zhang (2006).

Figure 5 gives another example, NOAA 11084, observed on July 2, 2010 of the solar cycle 24. As in Figure 4, the top two panels present the continuum image of the sunspot and the corresponding electric current distribution. Here the circles represent where $r$ is $5\arcsec$, $10\arcsec$ and $15\arcsec$ respectively. Similar trend as that in Figure 4 can be seen from the bottom panels of $h_{c}^1$ and $\alpha_{z}^1$. Here the average $h_{c}^1$ or $\alpha_{z}^1$ values change their sign at about $4\arcsec$. The mean value of $|B_z|$ in $r\leq5\arcsec$ is 2382 G, and is 713 G in $5\arcsec<r\leq20\arcsec$ region. The mean value of $\alpha_{z}^1$ in $r\leq5\arcsec$ is -1.300$\times10^{-8}m^{-1}$ and is 2.950$\times10^{-8}m^{-1}$ in $5\arcsec<r\leq20\arcsec$ region. The mean value of $h_{c}^1$ in $r\leq5\arcsec$ is -0.901 $\times 10^{-2}G^{2}m^{-1}$ and is 0.315$\times10^{-2}G^{2}m^{-1}$ in $5\arcsec<r\leq20\arcsec$. For the whole AR, $\alpha^1_{z}=3.599\times10^{-8}m^{-1}$ and $\alpha^1_{hc}=1.910\times10^{-8}m^{-1}$ with $|B_{z}| > $ 100 G. We see here again that the inner umbra and outer penumbra has the opposite helicity sign and the helicity sign of the whole AR is dominated by the sign in penumbra.

Note that Chatterjee et al. (2006) modeled the penetration of a poloidal field into a toroidal rising flux tube through turbulence diffusion. One important prediction of their model is the existence of a ring of reverse current helicity on the periphery of active regions. Our observations seem consistent with their prediction.

\section{Conclusion and Discussion}

Using high quality magnetograms taken with SP/Hinode we examined the hemispheric helicity sign rule in the descending phase of solar cycle 23 and the ascending phase of solar cycle 24. We studied two helicity parameters, $\alpha_{z}$ and $\alpha_{hc}$, of 64 actives regions, 30 belonging to solar cycle 23 and 34 belonging to solar cycle 24. We also examined how the hemispheric helicity sign rule depends on the selection of field points and whether strong and weak fields have opposite helicity sign as reported before.

Our analysis gives following results. (1) The 34 active regions in the ascending phase of the solar cycle 24 follow the so-called hemispheric helicity sign rule. (2) The 30 active regions in the descending phase of the solar cycle 23 do not follow the usual hemispheric helicity sign rule. (3) When combining all 64 active regions as one sample, the usual hemispheric helicity rule is indicated as in most other observations. (4) Even though we have used the so-far most accurate measurement of vector magnetic field given by SP/Hinode, the observed hemispheric helicity sign rule is still weak with large scatters. (5) The data show evidence of opposite helicity signs between strong and weak fields, and this is a presentation of that the helicity parameters change sign from the inner umbra to the outer penumbra.

We argue that results No. (1), (3) and (4) are consistent with the model by Longcope et al. (1998), result No. (5) is consistent with the model by Chatterjee et al. (2006). Results No. (1) and (2) seem suggesting that Choudhuri et al. (2004) has a merit in its physical picture, but may need to modify their result on which phase of the solar cycle that deviations from the hemispheric rule take place.  From our observations we speculate that both the $\Sigma$-effect (Longcope et al. 1998) and the dynamo (Choudhuri et al. 2004; Chatterjee et al. 2006) have contributed in the generation of helicity, whereas in both models turbulence in the convection zone has played an important role.


\acknowledgements

We thank the referee for helpful comments and suggestions. Hinode is a Japanese mission developed and launched by ISAS/JAXA, with NAOJ as domestic partner and NASA and STFC (UK) as international partners. It is operated by these agencies in co-operation with ESA and NSC (Norway). Hinode SOT/SP Inversions were conducted at NCAR under the framework of the Community Spectro-polarimetric Analysis Center (CSAC; http://www.csac.hao.ucar.edu/). This work was partly supported by the National Natural Science Foundation of China (Grant No. 10921303), Knowledge Innovation Program of Chinese Academy of Sciences (Grant No. KJCX2-EW-T07) and National Basic Research Program of MOST (Grant No. 2011CB811401).



\clearpage
\begin{figure}
\centering
\includegraphics[width=0.95\textwidth]{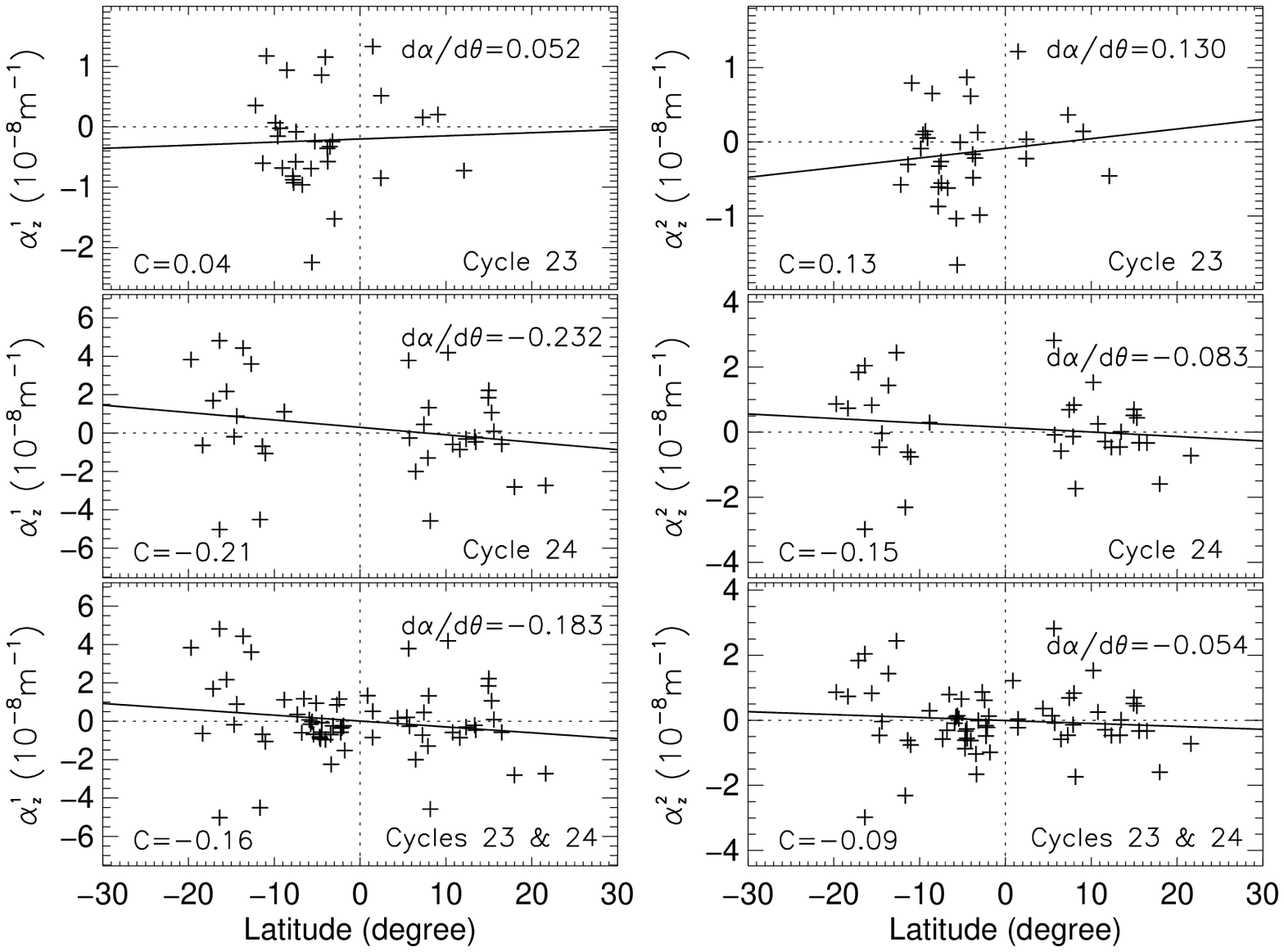}
\caption{\small Variation of $\alpha_{z}^1$ (left panels) and $\alpha_{z}^2$ (right panels) with the solar latitude for the 30 ARs in the descending phase of solar cycle 23 (top panels), the 34 ARs in the ascending phase of solar cycle 24 (middle panels) and the total 64 ARs (bottom panels). Here $\alpha_{z}^1$ and $\alpha_{z}^2$ are calculated using only points with $|B_{z}| > $ 100G. The solid lines indicate the results of least-square linear fits. Values of $d\alpha/d\theta$ from the linear fittings are shown in each panel, in the unit of $10^{-9}m^{-1}deg^{-1}$. Shown also in the left-bottom corner of each panel are the correlation coefficients between latitude and $\alpha_{z}^1$ or $\alpha_{z}^2$.}
\end{figure}

\begin{figure}
\centering
\includegraphics[width=0.95\textwidth]{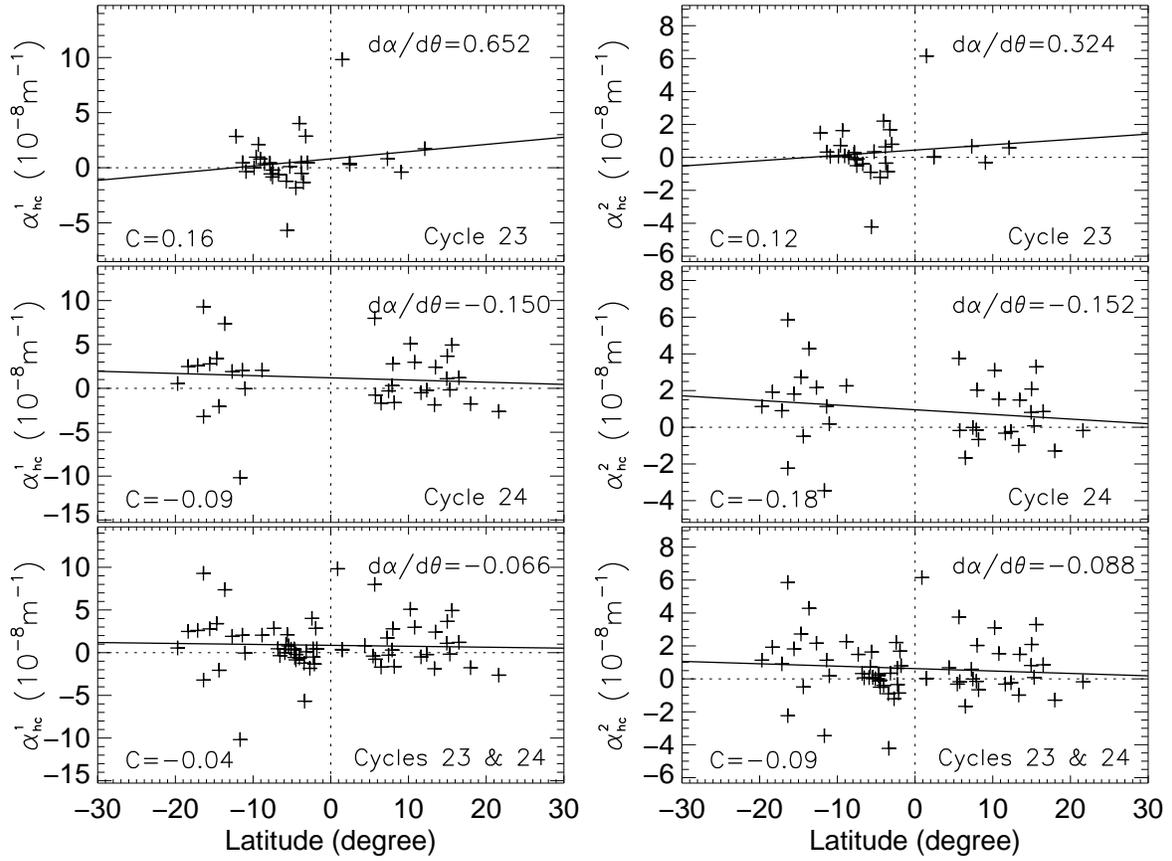}
\caption{\small Same as Figure 1, but for the parameters $\alpha_{hc}^1$ (left panels) and $\alpha_{hc}^2$ (right panels).}\mbox{}\\
\end{figure}

\begin{figure}
\centering
\includegraphics[width=0.95\textwidth]{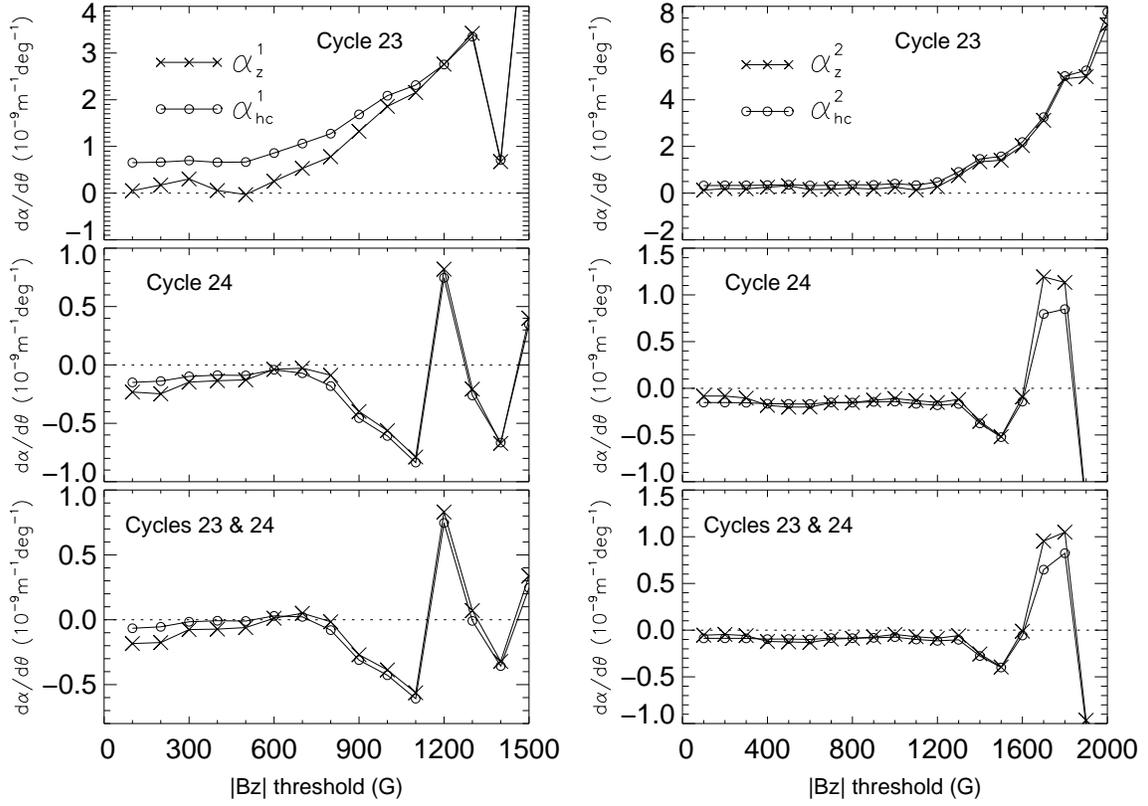}
\caption{\small Variation of the latitudinal gradient ($d\alpha/d\theta$) with different $|B_z|$ threshold (see text for details).  The cross symbols indicate the $d\alpha/d\theta$ values for $\alpha_{z}^1$ or $\alpha_{z}^2$, and the circles for $\alpha_{hc}^1$ or $\alpha_{hc}^2$.}\mbox{}\\
\end{figure}

\begin{figure}
\centering
\includegraphics[width=0.85\textwidth]{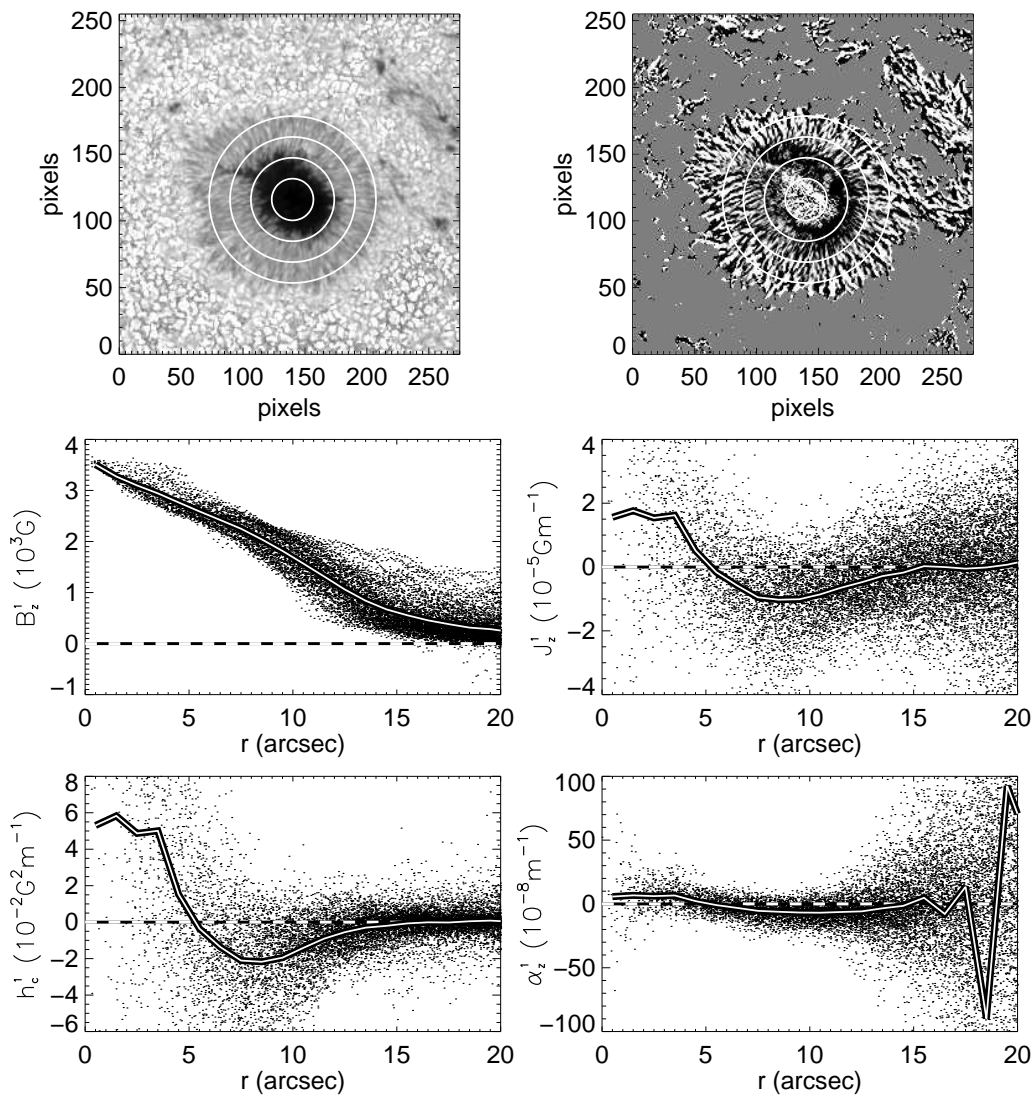}
\caption{\small The top left panel shows the continuum image of NOAA 10940 sunspot observed on Feb 1, 2007, belonging to the solar cycle 23. The $X$ and $Y$ spatial resolution of the image is 0.2971$''$/pixel and 0.3199$''$/pixel respectively. The top right panel shows corresponding electric current distribution. Circles in these two panels show where the distance to the sunspot center ($r$) are $5\arcsec$, $10\arcsec$, $15\arcsec$ and $20\arcsec$ respectively. The middle left panel shows the variation of $B_{z}^1$ with $r$. The dots show the values of $B_{z}^1$ and the double-shelled curve shows the mean value of $B_{z}^1$ with a bin of $1\arcsec$ in $r$. The middle right, bottom left and bottom right panels are similar to the middle left one, but for the values of electric current, $h_c^1$ and $\alpha_{z}^1$ respectively.}\mbox{}\\
\end{figure}

\begin{figure}
\centering
\includegraphics[width=0.85\textwidth]{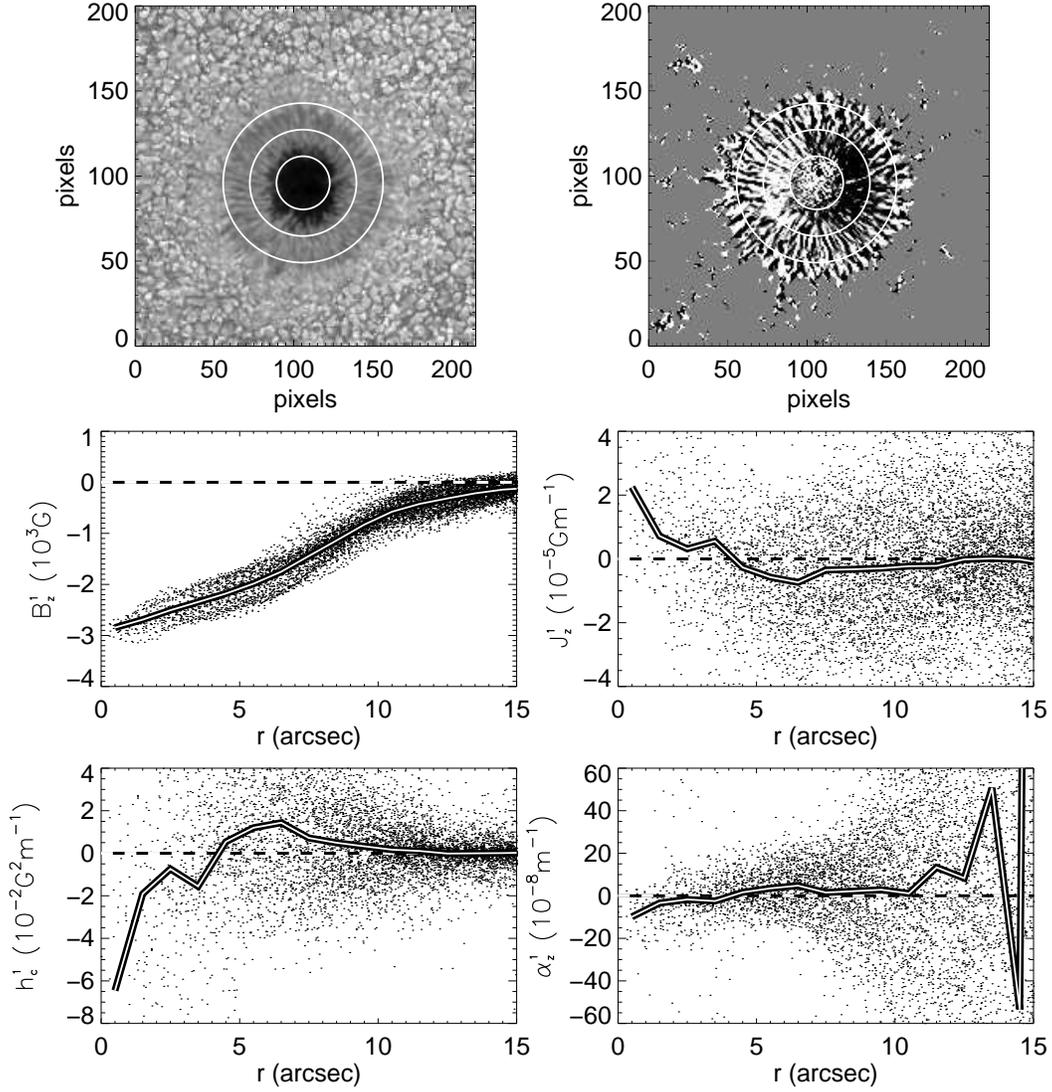}
\caption{\small Same as Figure 4, but for NOAA 11084 sunspot observed on July 2, 2010, belonging to the solar cycle 24. The spatial resolutions of the top images are the same to those in Figure 4, with the circles showing where $r$ is $5\arcsec$, $10\arcsec$ and $15\arcsec$ respectively.}\mbox{}\\
\end{figure}

\end{document}